\def\rfr#1{eq. (\ref{#1})}

\def\cf#1#2{\dot\Omega^{\rm #2}_{.#1}}

\def\dert#1#2{\frac{{{d}}{#1}}{{{d}}{#2}}}              

\def\bar{\begin{eqnarray}}
\def\ear{\end{eqnarray}}
\def\bb{\bibitem}
\def\eqi{\begin{equation}}
\def\eqf{\end{equation}}
\def\eqia{\begin{eqnarray}}
\def\eqfa{\end{eqnarray}}
\def\rp#1#2{{#1\over#2}}

\def\lb#1{\label{#1}}

\def\dr{\bds A_{\rm drag}}

\def\virg#1{``#1''}

\def\oc2{$\mathcal{O}(c^{-2})$}

\def\bds#1{\vec{\it{#1}}}

\documentclass{PoS}

\title{Recent Attempts to Measure the General Relativistic Lense-Thirring Effect with Natural and Artificial Bodies in the Solar System}

\ShortTitle{Attempts to Measure the Lense-Thirring Effect in the Solar System}

\author{\speaker{Lorenzo Iorio}\\
        INFN-Sezione di Pisa\\
        E-mail: \email{lorenzo.iorio@libero.it}}

\abstract{
According to general relativity, a spinning body of mass $M$ and angular momentum $\bds S$, like a star or a planet, generates a gravitomagnetic field which induces, among other phenomena, also the Lense-Thirring effect, i.e. secular precessions of the path of a test particle orbiting it. Direct and indisputable tests of such a relativistic prediction  are still missing.
We  discuss some performed attempts  to measure it in the gravitational fields of several  bodies in the Solar System with natural and artificial objects.
 The focus is on the realistic evaluation of the impact of some competing classical forces regarded  as sources of systematic uncertainties degrading the total accuracy obtainable.
In the case of the test performed with the LAGEOS and LAGEOS II Earth's satellites one of the major sources of systematic uncertainty is  the imperfect knowledge of the even zonal harmonic coefficients of the multipolar expansion of the Newtonian part of the terrestrial gravitational potential. The lingering uncertainty in some of them makes the total error in such a test as large as $15-30\%$, contrary to  more optimistic evaluations ($\approx 5-10\%$). Some different strategies to extract the relativistic signature from the data of the LAGEOS satellites which may complement and robustly corroborate the so far implemented tests are suggested. We critically discuss the possibilities that the LARES satellite, to be launched at the end of 2009, will realistically allow to measure the Lense-Thirring effect with a $\approx 1\%$ accuracy. Since it will orbit at much lower altitude than LAGEOS and LAGEOS II, much more even zonals will have to be accounted for; according to the present-day uncertainty in them, calculations performed with standard geodetic techniques show that their impact may be orders of magnitude larger than the expected accuracy level. Recent progresses in the planetary orbit determination make the perspective of reliably measuring the Sun's Lense-Thirring effect  a realistic possibility. Presently, the magnitude of the gravitomagnetic perihelion precessions of the inner planets is about of the same order of magnitude of the present-day uncertainties in determining the secular perihelion precessions from the observations of the rocky planets. Moreover, the predicted Lense-Thirring effect for all of them are in agreement with the estimated corrections to the standard Newtonian/Einsteinian perihelion precessions. Finally, we discuss a recent interpretation of some orbital data of the Mars Global Surveyor  spacecraft in terms of the gravitomagnetic field of Mars which recently raised a debate.}

\FullConference{5th International School on Field Theory and Gravitation,\\
		 April 20 - 24 2009\\
		 Cuiab$\acute{a}$ city, Brazil}

\begin{document}

\section{Introduction}
In the weak-field and slow motion approximation,  the Einstein field equations of the general theory of relativity, which is a  highly non-linear Lorentz-covariant theory of gravitation, get linearized resembling to the Maxwellian equations of electromagntism. As a consequence, a gravitomagnetic field $\bds B_{\rm g}$, induced by the off-diagonal components $g_{0i}, i=1,2,3$ of the space-time metric tensor related to the mass-energy currents of the source of the gravitational field, arises \cite{MashNOVA}. The gravitomagnetic field affects orbiting test particles, precessing gyroscopes, moving clocks and atoms and propagating electromagnetic waves \cite{Rug,Scia04}. Perhaps, the most famous gravitomagnetic effects are the spin-spin precession  of the axis of a gyroscope \cite{Pugh,Schi} and the spin-orbit Lense-Thirring\footnote{According to an interesting historical analysis recently performed by Pfister  \cite{Pfi07}, it would be more correct to speak about an Einstein-Thirring-Lense effect.} precessions \cite{LT} of the orbit of a test particle, both occurring in the field of a central slowly rotating mass like, e.g., our planet. For an analysis of the ambiguity arising from the use of the terminology frame-dragging, de Sitter precession\footnote{It is another precessional effect predicted by general relativity \cite{desi,Fok} which is not related to the mass-energy currents of the source.} and Lense-Thirring precession, in contrast to the unambiguous reference to spin-orbit and spin-spin precessions, see Ref.~\cite{Oco}.  Direct, undisputable measurements of such  fundamental predictions of general relativity are not yet available.

The measurement of the gyroscope precession in the Earth's gravitational field has been the goal of the dedicated space-based GP-B mission \cite{Eve,GPB} launched in 2004 and carrying onboard four superconducting gyroscopes; its data analysis is still ongoing. The target accuracy was originally $1\%$, but it is still unclear if the GP-B team will succeed in reaching such a goal because of some unmodelled effects affecting the gyroscopes: 1) a time variation in the polhode motion of the gyroscopes and 2) very large classical misalignment torques on the gyroscopes. For early results, see Ref.~\cite{primi}; see also on the WEB http://einstein.stanford.edu/.

\begin{figure}
   \caption{Keplerian orbit. The longitude of the ascending node $\Omega$ is counted from a reference X direction  in the equator of the central body, assumed as reference plane $\{{\rm X,Y}\}$,  to the line of the nodes which is the intersection of the orbital plane with the equatorial plane of the central body. It has mass $M$ and proper angular momentum $\bds S$. The argument of pericentre $\omega$ is an angle in the orbital plane counted from the line of the nodes to the location of the pericentre, here marked with $\Pi$. The time-dependent position of the moving test particle of mass $m$ is given by the true anomaly $f$, counted anticlockwise from the pericentre's position. The inclination between the orbital and the equatorial planes is $i$. Thus, $\Omega,\omega, i$ can be viewed as the three (constant) Euler angles fixing the configuration of a rigid body, i.e. the orbit which in the unperturbed Keplerian case does  change neither its shape nor its size, in the inertial $\{{\rm X,Y,Z}\}$ space.
   Courtesy by H.I.M. Lichtenegger, IWF, Graz.}
   \label{plo}
   \includegraphics{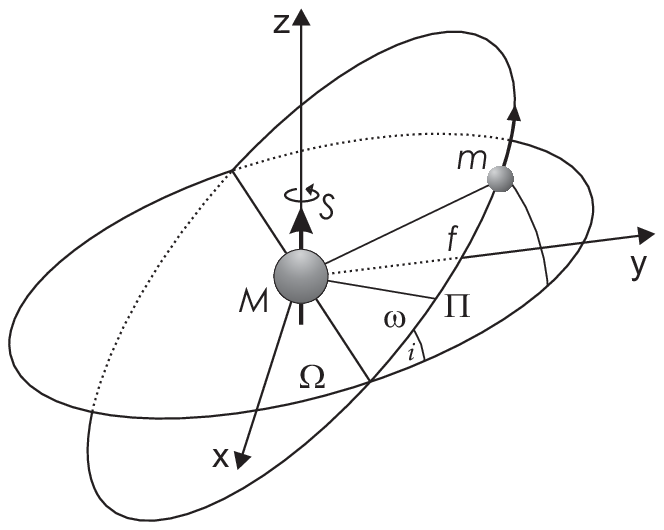}
   \end{figure}

In this Lecture we  will focus on the attempts to measure of the Lense-Thirring effect in the gravitational fields of the Earth, the Sun and Mars. Far from a localized rotating body  with angular momentum ${\bds S}$  the gravitomagnetic field can be written as
\eqi \bds B_{\rm g} = -\rp{G}{c r^3}\left[\bds S -3\left(\bds S\cdot{\hat{r}}\right){\hat{r}}\right],\eqf   where $G$ is the Newtonian gravitational constant and $c$ is the speed of light in vacuum. It acts on a test particle orbiting with a velocity $\bds v$ with the non-central acceleration \cite{Sof}
\eqi \bds A_{\rm LT} = -\rp{2}{c}{\bds v}\times\bds B_{\rm g}\eqf
which induces secular precessions of the longitude of the ascending node $\Omega$ \begin{equation}\dot\Omega_{\rm LT} = \rp{2 G S}{c^2 a^3 (1-e^2)^{3/2}},\lb{let}\end{equation}
and the argument of pericentre $\omega$
\begin{equation}\dot\omega_{\rm LT} = -\rp{6 G S\cos i}{c^2 a^3 (1-e^2)^{3/2}},\lb{LT_o}\end{equation}
of the orbit of  a test particle. In \rfr{let} and \rfr{LT_o}
 $a$ and $e$ are the semimajor axis and the eccentricity, respectively, of the test particle's orbit and $i$ is its inclination to the central body's equator.
The semimajor axis $a$ fixes the size of the ellipse, while its shape is determined by the eccentricity $0\leq e<1$; an orbit with $e=0$ is a circle. The angles $\Omega$ and $\omega$ establish the orientation of the orbit in the inertial space and in the orbital plane, respectively. $\Omega$, $\omega$ and $i$ can be viewed as the three Euler angles which determine the orientation of a rigid body with respect to an inertial frame. In Figure \ref{plo} we illustrate the geometry of a Keplerian orbit \cite{Roy}.

In this Lecture we will critically discuss the following  topics
\begin{itemize}
  \item Section \ref{grav}. \emph{The realistic evaluation  of the total accuracy in the test performed in recent years with the existing Earth's artificial satellites
  LAGEOS and LAGEOS II \cite{Ciu04,Ciu06,Ries08}}.

  LAGEOS was put into orbit in 1976, followed by its twin LAGEOS II in 1992; they are passive, spherical spacecraft entirely covered by retroreflectors which allow for their accurate tracking through laser pulses sent from Earth-based ground stations according to the Satellite Laser Ranging (SLR) technique \cite{slr}. They orbit at altitudes of about 6000 km ($a_{\rm LAGEOS} =12270$ km, $a_{\rm LAGEOS\ II}=12163$ km) in nearly circular paths ($e_{\rm LAGEOS}=0.0045$, $e_{\rm LAGEOS\ II}=0.014$) inclined by 110 deg and 52.65 deg, respectively, to the Earth's equator. The Lense-Thirring effect for their nodes amounts to about 30 milliarcseconds per year (mas yr$^{-1}$) which correspond to about 1.7 m yr$^{-1}$ in the cross-track direction\footnote{A perturbing acceleration like $\bds A_{\rm LT}$ is customarily projected onto the radial ${\hat r}$, transverse ${\hat \tau}$ and cross-track ${\hat \nu}$ directions of an orthogonal  frame comoving with the satellite \cite{Sof}; it turns out that the Lense-Thirring node precession affects the cross-track component of the orbit according to $\Delta\nu_{\rm LT} \approx a\sin i\Delta\Omega_{\rm LT}$ (eq. (A65), p. 6233 in Ref.~\cite{Cri}).} at the LAGEOS altitudes.

  The idea of measuring the Lense-Thirring node rate with the just launched LAGEOS satellite, along with the other SLR targets orbiting at that time, was put forth by Cugusi and Proverbio \cite{Cug2,Cug78}. Tests have started to be effectively performed later by using the LAGEOS and LAGEOS II satellites \cite{tanti}, according to a strategy  by Ciufolini \cite{Ciu96} involving the use of a suitable linear combination of the nodes $\Omega$ of both satellites and the perigee $\omega$ of LAGEOS II. This was done to reduce the impact of the most relevant source of systematic bias, i.e. the mismodelling in the even ($\ell=2,4,6\ldots$) zonal ($m=0$) harmonic coefficients $J_{\ell}$ of the multipolar expansion of the Newtonian part of the terrestrial gravitational potential due to the diurnal rotation (they induce secular precessions on the node and perigee of a terrestrial satellite much larger than the gravitomagnetic ones. The $J_{\ell}$ coefficients cannot be theoretically computed but must be estimated by suitably processing long data sets from the dedicated satellites like CHAMP and GRACE; see Section \ref{grav}): the three-elements combination used allowed for removing the uncertainties in $J_2$ and $J_4$. In Ref.~\cite{Ciu98} a $\approx 20\%$ test was reported by using the\footnote{Contrary to the subsequent models based on the dedicated satellites CHAMP (http://www-app2.gfz-potsdam.de/pb1/op/champ/index$\_$CHAMP.html) and GRACE (http://www-app2.gfz-potsdam.de/pb1/op/grace/index$\_$GRACE.html), EGM96 relies upon multidecadal tracking of SLR data of a constellation of geodetic satellites including LAGEOS and LAGEOS II as well; thus the possibility of a sort of $a-priori$ `imprinting' of the Lense-Thirring effect itself, not solved-for in EGM96, cannot be neglected.} EGM96 Earth gravity model \cite{Lem98}; subsequent detailed analyses showed that such an evaluation of the total error budget was overly optimistic in view of the likely unreliable computation of the total bias due to the even zonals \cite{Ior03,Ries03a,Ries03b}.  An analogous, huge underestimation turned out to hold also for the effect of the non-gravitational perturbations \cite{Mil87} like the direct solar radiation pressure, the Earth's albedo, various subtle thermal effects depending on  the physical properties of the satellites' surfaces and their rotational state \cite{Inv94,Ves99,Luc01,Luc02,Luc03,Luc04,Lucetal04,Ries03a}, which the perigees  of LAGEOS-like satellites are particularly sensitive to. As a result, the realistic total error budget in the test reported in \cite{Ciu98} might be as large as $60-90\%$ or (by considering EGM96 only) even more.

The observable used\footnote{The first test was, in fact, performed with the CHAMP-only based model EIGEN-2 \cite{EIGEN2} by Lucchesi in Ref.~\cite{Luccprimo}.} in Ref.~\cite{Ciu04} with the GRACE-only EIGEN-GRACE02S model \cite{eigengrace02s} and in Ref.~\cite{Ries08} with other more recent Earth gravity models was the following linear combination\footnote{See also Refs.~\cite{Pav02,Ries03a,Ries03b}.} of the nodes of LAGEOS and LAGEOS II, explicitly computed in Ref.~\cite{IorMor} following the approach put forth in Ref.~\cite{Ciu96}
\begin{equation} f=\dot\Omega^{\rm LAGEOS}+
c_1\dot\Omega^{\rm LAGEOS\ II }, \lb{combi}\end{equation}
where \begin{equation} c_1\equiv-\rp{\dot\Omega^{\rm LAGEOS}_{.2}}{\dot\Omega^{\rm
LAGEOS\ II }_{.2}}=-\rp{\cos i_{\rm LAGEOS}}{\cos i_{\rm LAGEOS\
II}}\left(\rp{1-e^2_{\rm LAGEOS\ II}}{1-e^2_{\rm
LAGEOS}}\right)^2\left(\rp{a_{\rm LAGEOS\ II}}{a_{\rm LAGEOS}}\right)^{7/2}.\lb{coff}\end{equation} The coefficients $\dot\Omega_{.\ell}$ of the aliasing classical node precessions \cite{Kau} $\dot\Omega_{\rm class}=\sum_{\ell}\dot\Omega_{.\ell}J_{\ell}$ induced by the even zonals  have been analytically worked out up to $\ell=20$ in, e.g., Ref.~\cite{Ior03}; they yield $c_1=0.544$.
 The Lense-Thirring signature of \rfr{combi} amounts to 47.8 mas yr$^{-1}$. The combination  \rfr{combi} allows, by construction, to remove the aliasing effects due to the static and time-varying parts of the first even zonal $J_2$. The nominal (i.e. computed with the estimated values of $J_{\ell}$, $\ell=4,6...$) bias due to the remaining higher degree even zonals would amount to  about $10^5$ mas yr$^{-1}$; the need of a careful and reliable modeling of such an important source of systematic bias is, thus, quite apparent. Conversely, the nodes of the LAGEOS-type spacecraft are directly affected by the non-gravitational accelerations at a $\approx 1\%$ level of the Lense-Thirring effect \cite{Luc01,Luc02,Luc03,Luc04,Lucetal04}. For a comprehensive, up-to-date overview of the numerous and subtle issues concerning the measurement of the Lense-Thirring effect see, e.g., Ref.~\cite{IorNOVA}.
\item Section \ref{approach}. \emph{Another approach which could be followed in extracting the Lense-Thirring effect from the data of the LAGEOS-type satellites}.
  \item Section \ref{larez}. \emph{The possibility that the LARES mission, recently approved by the Italian Space Agency (ASI), will be able to measure the Lense-Thirring node precession with an accuracy of the order of $1\%$}.

      In Refs.~\cite{vpe76a,vpe76b}  it was proposed to measure the Lense-Thirring precession of the nodes $\Omega$ of a pair of counter-orbiting spacecraft to be launched in terrestrial polar orbits and endowed with drag-free apparatus. A somewhat equivalent, cheaper version of such an idea was put forth in 1986 by Ciufolini in Ref.~\cite{Ciu86} who proposed to launch a passive, geodetic satellite in an orbit identical to that of LAGEOS  apart from the orbital planes which should have been displaced by 180 deg apart. The measurable quantity was, in the case of the proposal of Ref.~\cite{Ciu86}, the sum of the nodes of LAGEOS and of the new spacecraft, later named LAGEOS III, LARES, WEBER-SAT, in order to cancel to a high level of accuracy the corrupting effect of the multipoles of the Newtonian part of the terrestrial gravitational potential which represent the major source of systematic error (see Section \ref{grav}). Although extensively studied by various groups \cite{CSR,LARES}, such an idea was not implemented for many years.
In Ref.~\cite{Ioretal02} it was proposed to include also the data from LAGEOS II by using a different observable.  Such an approach was proven in Ref.~\cite{IorNA} to be potentially useful in making the constraints on the orbital configuration of the new SLR satellite less stringent than it was originally required in view of the recent improvements in our knowledge of the classical part of the terrestrial gravitational potential due to the dedicated CHAMP and, especially, GRACE  missions.

Since reaching high altitudes and minimizing the unavoidable orbital injection errors is expensive, it was explored the possibility of discarding LAGEOS and LAGEOS II using a low-altitude, nearly polar orbit for LARES \cite{LucPao01,Ciu06b}, but in Refs.~\cite{Ior02,Ior07c} it was proven that such alternative approaches are not feasible. It was also suggested that LARES would be able to probe alternative theories of gravity \cite{Ciu04b}, but also in this case it turned out to be impossible \cite{IorJCAP,Ior07d}.

The stalemate came to an end when about one year ago ASI  made the following official announcement\footnote{It was posted on the currently expired webpage http://www.asi.it/SiteEN/MotorSearchFullText.aspx?keyw=LARES.}: ``On February 8, the ASI board approved funding for the LARES mission, that will be launched with VEGA's maiden flight before the end of 2008. LARES is a passive satellite with laser mirrors, and will be used to measure the Lense-Thirring effect.''  The italian version of the announcement yields some more information specifying that LARES, designed in collaboration with National Institute of Nuclear Physics (INFN), is currently under construction by Carlo Gavazzi Space SpA; its Principal Investigator (PI) is I. Ciufolini and its scientific goal is to measure at a $1\%$ level the  Lense-Thirring effect in the gravitational field of the Earth.
Concerning the orbital configuration of LARES,
in one of the latest communication to INFN, Rome, 30 January 2008, Ciufolini \cite{INFN} writes that LARES will be launched with a semimajor axis of approximately 7600 km and an inclination between 60 and 80 deg.
More precise information can be retrieved in Section 5.1, pag 9 of the document Educational Payload on the Vega Maiden Flight
Call For CubeSat Proposals, European Space Agency,
Issue 1
11 February 2008, downloadable at
http://esamultimedia.esa.int/docs/LEX-EC/CubeSat$\%$20CFP$\%$20issue$\%$201.pdf.
    It is  written there that LARES will be launched into a circular orbit with altitude $h=1200$ km, corresponding to a semimajor axis $a_{\rm LARES}=7578$ km, and inclination $i=71$ deg to the Earth's equator. Latest information\footnote{See on the WEB
http://www.esa.int/esapub/bulletin/bulletin135/bul135f$\_$bianchi.pdf.} point towards a launch at the end of 2009 with a VEGA rocket in a circular orbit inclined by 71 deg to the Earth's equator at an altitude of\footnote{I thank Dr. D. Barbagallo (ESRIN) for having kindly provided me with the latest details of the orbital configuration of LARES.} 1450 km corresponding to a semimajor axis of $a_{\rm LR}=7828$ km. More or less the same has been reported by Ciufolini  to the INFN in Villa Mondragone, 3 October 2008 \cite{INFN2}, and to the delegates of the 16th International Workshop on Laser Ranging, 13-17 October 2008 \cite{CiufoILRS}.
However, according to more recent information from the official ASI website (http://www.asi.it/en/activity/cosmology/lares), it seems that there is still a broad uncertainty in the final orbital configuration of LARES which should, now, correspond to $a\approx 7778$ km and $60$ deg $< i < 86$ deg.
\item  Section \ref{SolLT} \emph{The possibility of measuring the Lense-Thirring effect of the Sun with the inner planets of the Solar System}.

Recent determinations of the Sun's proper angular
momentum \eqi S_{\odot}=(190.0\pm 1.5)\times 10^{39}\ {\rm kg\ m}^2\ {\rm s}^{-1}\eqf from helioseismology
\cite{Pij1,Pij2}, accurate to $0.8\%$, yields a value about one order of magnitude smaller than that  obtained by assuming an homogeneous and uniformly rotating  Sun, as done in the pionieristic work by De Sitter  \cite{DeS}, and also in Refs.~\cite{Sof,Cug78} who concluded that, at their time, it was not possible to measure the solar Lense-Thirring effect. Now, instead, the situation seems much more promising, in spite of the reduced magnitude of the solar gravitomagnetic field with respect to the earlier predictions.
As in the Schwarzschild metric the characteristic length to be compared with the accuracy in determining particles'orbits is the Schwarzschild radius $R_g=2GM/c^2=3$ km for the Sun, the  typical solar gravitomagnetic length is \eqi l^{\odot}_g = \rp{S_{\odot}}{M_{\odot}c} =319\ {\rm m}.\eqf The present-day accuracy in knowing, e.g., the inner planet's mean radius  \eqi\left\langle r\right\rangle=a\left(1+\rp{e^2}{2}\right),\eqf is shown in Table \ref{chebol}.
\begin{table}
\caption{First line: uncertainties, in m, in the average heliocentric distances of the inner planets obtained by propagating the formal errors in $a$ and $e$ according to Table 3 of Ref.~\protect\cite{Pit08}; the EPM2006 ephemerides were used by E.V. Pitjeva \protect\cite{Pit08}. Second line: maximum differences, in m, between the EPM2006 and the DE414 \protect\cite{DE414} ephemerides for thr inner planets in the time interval 1960-2020 according to Table 5 of Ref.~\protect\cite{PROO}. They have to be compared to the characteristic gravitomagnetic length of the Sun $l_g^{\odot}=319$ m.\label{chebol}
}
\centering
\bigskip
\begin{tabular}{@{}lllll@{}}
\noalign{\smallskip}\hline\noalign{\smallskip}
Type of orbit uncertainty & {Mercury} & {Venus} & {Earth} & {Mars}  \\
 \noalign{\smallskip}\hline\noalign{\smallskip}
$\delta\left\langle r \right\rangle$ (EPM2006)  & 38 & 3 & 1 & 2 \\
EPM2006$-$DE414 & 256 & 131 & 17.2 & 78.7 \\
 \noalign{\smallskip}\hline\noalign{\smallskip}

\end{tabular}
\end{table}
Such values have been obtained by linearly propagating the formal, statistical errors in $a$ and $e$ according to Table 3 of Ref.~\cite{Pit08}; it is clear that, even by re-scaling them by a a factor of, say, $2-5$, the gravitomagnetic effects due to the Sun's rotation fall in the measurability domain. Another possible way to evaluate the present-day uncertainty in the planetary orbital motions consists of looking at different ephemerides of comparable accuracy. In Table \ref{chebol} we do that for the EPM2006 \cite{Pit08} and the DE414 \cite{DE414} ephemerides; although, larger than $\delta\left\langle r\right\rangle$, the maximum differences between such ephemerides are smaller than the solar gravitomagnetic length $l_g^{\odot}$.
Since much ranging data to Venus Express spacecraft\footnote{See on the WEB http://www.esa.int/SPECIALS/Venus$\_$Express/.} will be collected and processed, and in view of the ongoing Messenger \cite{Balo} and the future Bepi-Colombo \cite{Bepi,Balo} missions to Mercury and of the developments in the Planetary Laser Ranging (PLR) technique \cite{PLR}, it seems that the planetary orbit accuracy will be further improved. It is
remarkable to note that the currently available estimate of
$S_{\odot}$	from helioseismology is accurate enough to allow, in principle, a genuine Lense-Thirring test. Moreover, it was determined in a relativity-free
fashion from astrophysical techniques which do not rely on
the dynamics of planets in the gravitational field of the
Sun. Thus, there is no any a priori \virg{memory} effect of general relativity itself  in the adopted value of $S_{\odot}$.

\item    Section \ref{MarsLT}. \emph{The test with the Mars Global Surveyor probe in the gravitational field of Mars}.

 Since the angular momentum of Mars can be evaluated as
 \eqi S_{\rm M}=(1.92\pm 0.01)\times 10^{32}\ {\rm kg\ m}^2\ {\rm s}^{-1}\eqf from the latest spacecraft-based determinations of the areophysical parameters quoted in Ref.~\cite{Kon},
 it turns out
 \eqi l^{\rm M}_g = \rp{S_{\rm M}}{M_{\rm M}c} =1.0\ {\rm m}.\eqf Such a value has to be compared with the present-day accuracy in determining the orbit of a spacecraft like, e.g.,  {Mars Global Surveyor}
(MGS); it can be evaluated it is about  $0.15$ m \cite{Kon} in the radial direction, not affected by the gravitomagnetic force itself. Thus, it makes sense to investigate the possibility of measuring the Lense-Thirring effect in the gravitational field of Mars as well. In fact,
 the {time series} of the {Root Mean Square} (RMS)
{orbit overlap differences} \cite{Kon} of the {out-of-plane} part
${\nu}$ of the orbit of the nearly polar MGS probe ($a=3796$ km, $i=92.86$ deg, $e=0.0085$) over a time span ${\Delta P\approx 5}$ {yr}, processed {without modelling gravitomagnetism} at all, has been successfully interpreted in terms of the Lense-Thirring effect by Iorio in Refs.~\cite{IorMGS,IorMGS2}. However, {criticisms} have been raised by Krogh in Ref.~\cite{Krogh}; Iorio has replied in Ref.~\cite{CEJP1}. In Section \ref{MarsLT} we will discuss them in details.

\end{itemize}

\section{The systematic error of gravitational origin in the LAGEOS-LAGEOS II test}\lb{grav}
The realistic evaluation of the total error budget of the LAGEOS-LAGEOS II node test \cite{Ciu04} raised a lively debate
\cite{Ciu05,Ciu06,IorNA,IorJoG,IorGRG,Ior07,Luc05,CEJP2}, mainly focussed on the impact of the static and time-varying parts of the Newtonian component of the Earth's gravitational potential through the secular precessions induced on a satellite's node.

In the real world the path of a probe is not only affected by the relativistic gravitomagentic field but also by a huge number of other competing classical orbital perturbations of gravitational and non-gravitational origin.
The most insidious disturbances are those induced by the static part of the Newtonian component of the multipolar expansion in spherical harmonics\footnote{The relation among the  even zonals $J_{\ell}$ and the  normalized gravity coefficients $\overline{C}_{\ell 0}$ which are customarily determined in the Earth's gravity models, is $J_{\ell}=-\sqrt{2\ell + 1}\ \overline{C}_{\ell 0}$.} $J_{\ell}, \ell = 2,4,6,...$ of the gravitational potential of the central rotating mass \cite{Kau}: they affect the node with effects having the same signature of the relativistic signal of interest, i.e. linear trends which are orders of magnitude larger and cannot be removed from the time series of data without affecting the Lense-Thirring pattern itself as well. The only thing that can be done is to model such a corrupting effect as most accurately as possible and assessing the impact of the residual mismodelling on the measurement of the frame-dragging effect.
The secular precessions induced by the even zonals of the geopotential can be written as
\begin{equation}\dot\Omega^{\rm geopot}=\sum_{\ell  =2}\dot\Omega_{.\ell}J_{\ell},\end{equation}
where the coefficients $\dot\Omega_{.\ell}, \ell=2,4,6,...$ depend on the parameters of the Earth ($GM$ and the equatorial radius $R$) and on the semimajor axis $a$, the inclination $i$ and the eccentricity $e$ of the satellite. For example, for $\ell=2$
we have
\begin{equation}\dot\Omega_{.2}=-\rp{3}{2}n\left(\rp{R}{a}\right)^2\rp{\cos i}{(1-e^2)^2};\end{equation}
 $n=\sqrt{GM/a^3}$ is the Keplerian mean motion.
They have been analytically computed up to $\ell=20$  in, e.g., \cite{Ior03}.
Their mismodelling can be written as
\begin{equation}\delta\dot\Omega^{\rm geopot}\leq \sum_{\ell  =2}\left|\dot\Omega_{.\ell}\right|\delta J_{\ell},\lb{mimo}\end{equation}
where $\delta J_{\ell}$ represents our uncertainty in the knowledge of the even zonals $J_{\ell}$

A common feature of all the competing evaluations  so far published is that the systematic bias due to the static component of  the
  geopotential was always calculated  by using the released (more or less accurately calibrated) sigmas $\sigma_{J_{\ell}}$  of one Earth gravity model solution at a time for the uncertainties $\delta J_{\ell}$. Thus, it was said that the model X yields a $x\%$ error, the model Y   yields a $y\%$ error, and so on.

Since a trustable calibration of the formal, statistical uncertainties in the estimated zonals of the covariance matrix of a global solution is always a difficult task to be implemented in a reliable way, a much more realistic and conservative approach consists, instead, of taking the difference\footnote{See Fig.5 of \cite{Luc07} for a comparison of the estimated $\overline{C}_{40}$ in different models.} \eqi\Delta J_{\ell}=\left|J_{\ell}(\rm X) - J_{\ell}(\rm Y)\right|,\ \ell=2,4,6,...\eqf of the estimated even zonals for different pairs of Earth gravity field solutions as representative of the real uncertainty $\delta J_{\ell}$ in the zonals \cite{Lerch}. In Table \ref{tavola1}--Table \ref{tavolaAIUB2}  we present our results for the most recent GRACE-based models released so far by different institutions and retrievable on the Internet at\footnote{I thank M Watkins (JPL) for having provided me with the even zonals and their sigmas of the  JEM01-RL03B model.}
http://icgem.gfz-potsdam.de/ICGEM/ICGEM.html.   The models used are EIGEN-GRACE02S \cite{eigengrace02s}  from GFZ (Potsdam, Germany), GGM02S \cite{ggm02} and GGM03S \cite{ggm03} from CSR (Austin, Texas), ITG-Grace02s \cite{ITG} and ITG-Grace03s
\cite{itggrace03s} from IGG (Bonn, Germany), JEM01-RL03B from JPL (NASA, USA) and AIUB-GRACE01S \cite{aiub} from AIUB (Switzerland).
Note that this approach was explicitly followed also by Ciufolini in Ref.~\cite{Ciu96} with the JGM3 and GEMT-2 models.
In Table \ref{tavola1}--Table \ref{tavolaAIUB2} we quote both the sum $\sum_{\ell=4}^{20}f_{\ell}$ of the absolute values of the individual mismodelled terms \eqi f_{\ell} =\left|\dot\Omega_{.\ell}^{\rm LAGEOS} + c_1\dot\Omega_{.\ell}^{\rm LAGEOS\ II}\right|\Delta J_{\ell}\eqf (SAV), and the square root of the sum of their squares $\sqrt{\sum_{\ell=4}^{20}f^2_{\ell}}$ (RSS); in both cases we normalized them to the combined Lense-Thirring total precession of 47.8 mas yr$^{-1}$.
\begin{table}[!h]
\caption{Impact of the mismodelling in the even zonal harmonics on
$f_{\ell}=\left|\dot\Omega^{\rm LAGEOS}_{\ell} + c_1\dot\Omega^{\rm LAGEOS\ II}_{.\ell}\right|\Delta J_{\ell},\ \ell=4,\dots,20$,
 in mas yr$^{-1}$. Recall that
$J_{\ell}=-\sqrt{2\ell + 1}\ \overline{C}_{\ell 0}$;
for the uncertainty in the even zonals we have taken here the difference
$\Delta\overline{C}_{\ell 0}=\left|\overline{C}_{\ell 0}^{\rm(X)}-\overline{C}_{\ell 0}^{\rm (Y)}\right|$
 between the model X=GGM02S \protect\cite{ggm02} and the model Y=ITG-Grace02s\protect\cite{ITG}.
GGM02S is based on 363 days of GRACE-only data   (GPS and intersatellite tracking, neither constraints nor regularization applied)
spread between April 4, 2002 and Dec 31, 2003. The $\sigma$ are formal for both models. $\Delta \overline{C}_{\ell 0}$ are always larger than the linearly added sigmas, apart from   $\ell=12$ and $\ell=18$. Values of $f_{\ell}$ smaller than 0.1 mas yr$^{-1}$ have not been quoted. The
Lense-Thirring precession of the combination of \protect\rfr{combi} amounts to 47.8 mas yr$^{-1}$. The percent bias $\delta\mu$ have been computed by
normalizing the linear sum of $f_{\ell}, \ell=4,\dots,20$ (SAV) and the square root of the sum of $f_\ell^2, \ell=4,\dots,20$ to the Lense-Thirring combined precessions.
}\label{tavola1}
\centering
\bigskip

\begin{tabular}{llll}
\hline\noalign{\smallskip}
$\ell$ & $\Delta\overline{C}_{\ell 0}$ (GGM02S-ITG-Grace02s) & $\sigma_{\rm  X}+\sigma_{\rm Y}$ & $f_{\ell}$  (mas yr$^{-1}$)\\
\noalign{\smallskip}\hline\noalign{\smallskip}
4 & $1.9\times 10^{-11}$ &  $8.7\times 10^{-12}$ & 7.2\\
6 & $2.1\times 10^{-11}$ &  $4.6\times 10^{-12}$ & 4.6\\
8 & $5.7\times 10^{-12}$ &  $2.8\times 10^{-12}$ & 0.2\\
10 & $4.5\times 10^{-12}$ &  $2.0\times 10^{-12}$ & -\\
12 & $1.5\times 10^{-12}$ &  $1.8\times 10^{-12}$ & -\\
14 & $6.6\times 10^{-12}$ &  $1.6\times 10^{-12}$ & -\\
16 & $2.9\times 10^{-12}$ &  $1.6\times 10^{-12}$ & -\\
18 & $1.4\times 10^{-12}$ &  $1.6\times 10^{-12}$ & -\\
20 & $2.0\times 10^{-12}$ &  $1.6\times 10^{-12}$ & -\\

\noalign{\smallskip}\hline\noalign{\smallskip}
 &    $\delta\mu = 25\%$ (SAV) & $\delta\mu = 18\%$ (RSS) &   \\  %
\noalign{\smallskip}\hline\noalign{\smallskip} %
\end{tabular}
\end{table}
%
%
%
%
%
%
%
%
%
%
%
%
%
%
%
%
%
%
%

\begin{table}[!h]
   \caption{Bias due to the mismodelling in the even zonals of the models X=ITG-Grace03s \protect\cite{itggrace03s}, based on GRACE-only accumulated normal equations from data out of September 2002-April 2007 (neither apriori information nor regularization used), and Y=GGM02S \protect\cite{ggm02}.  The $\sigma$ for both models are formal. $\Delta \overline{C}_{\ell 0}$ are always larger than the linearly added sigmas, apart from  $\ell=12$ and $\ell=18$.}\label{tavola11}
\centering
\bigskip

\begin{tabular}{llll}
\hline\noalign{\smallskip}
$\ell$ & $\Delta\overline{C}_{\ell 0}$ (ITG-Grace03s-GGM02S) & $\sigma_{\rm  X}+\sigma_{\rm Y}$ & $f_{\ell}$  (mas yr$^{-1}$)\\
\noalign{\smallskip}\hline\noalign{\smallskip}
4 & $2.58\times 10^{-11}$ &  $8.6\times 10^{-12}$ & 9.6\\
6 & $1.39\times 10^{-11}$ &  $4.7\times 10^{-12}$ & 3.1\\
8 & $5.6\times 10^{-12}$ &  $2.9\times 10^{-12}$ & 0.2\\
10 & $1.03\times 10^{-11}$ &  $2\times 10^{-12}$ & -\\
12 & $7\times 10^{-13}$ &  $1.8\times 10^{-12}$ & -\\
14 & $7.3\times 10^{-12}$ &  $1.6\times 10^{-12}$ & -\\
16 & $2.6\times 10^{-12}$ &  $1.6\times 10^{-12}$ & -\\
18 & $8\times 10^{-13}$ &  $1.6\times 10^{-12}$ & -\\
20 & $2.4\times 10^{-12}$ &  $1.6\times 10^{-12}$ & -\\

\noalign{\smallskip}\hline\noalign{\smallskip}
&    $\delta\mu = 27\%$ (SAV) & $\delta\mu = 21\%$ (RSS) &   \\  %
\noalign{\smallskip}\hline\noalign{\smallskip} %
\end{tabular}
\end{table}
\begin{table}[!h]
   \caption{Bias due to the mismodelling in the even zonals of the models  X = GGM02S \protect\cite{ggm02} and Y = GGM03S \protect\cite{ggm03} retrieved from data spanning January 2003 to December 2006.
    The $\sigma$ for GGM03S are calibrated. $\Delta \overline{C}_{\ell 0}$ are larger than the linearly added sigmas for $\ell = 4,6$. (The other zonals are of no concern)}\label{tavola03S}
\centering
\bigskip

\begin{tabular}{llll}
\hline\noalign{\smallskip}
$\ell$ & $\Delta\overline{C}_{\ell 0}$ (GGM02S-GGM03S) & $\sigma_{\rm  X}+\sigma_{\rm Y}$ & $f_{\ell}$  (mas yr$^{-1}$)\\
\noalign{\smallskip}\hline\noalign{\smallskip}
4 & $1.87\times 10^{-11}$ &  $1.25\times 10^{-11}$ & 6.9\\
6 & $1.96\times 10^{-11}$ &  $6.7\times 10^{-12}$ & 4.2\\
8 & $3.8\times 10^{-12}$ &  $4.3\times 10^{-12}$ & 0.1\\
10 & $8.9\times 10^{-12}$ &  $2.8\times 10^{-12}$ & 0.1\\
12 & $6\times 10^{-13}$ &  $2.4\times 10^{-12}$ & -\\
14 & $6.6\times 10^{-12}$ &  $2.1\times 10^{-12}$ & -\\
16 & $2.1\times 10^{-12}$ &  $2.0\times 10^{-12}$ & -\\
18 & $1.8\times 10^{-12}$ &  $2.0\times 10^{-12}$ & -\\
20 & $2.2\times 10^{-12}$ &  $1.9\times 10^{-12}$ & -\\

\noalign{\smallskip}\hline\noalign{\smallskip}
&    $\delta\mu = 24\%$ (SAV) & $\delta\mu = 17\%$ (RSS) &   \\  %
\noalign{\smallskip}\hline\noalign{\smallskip} %
\end{tabular}
\end{table}
\begin{table}[!h]
   \caption{Bias due to the mismodelling in the even zonals of the models  X = EIGEN-GRACE02S \protect\cite{eigengrace02s} and Y = GGM03S \protect\cite{ggm03}.
    The $\sigma$ for both models are calibrated. $\Delta \overline{C}_{\ell 0}$ are always larger than the linearly added sigmas apart from $\ell = 14,18$.}\label{tavola033S}
\centering
\bigskip

 \begin{tabular}{llll}
\hline\noalign{\smallskip}
$\ell$ & $\Delta\overline{C}_{\ell 0}$ (EIGEN-GRACE02S-GGM03S) & $\sigma_{\rm  X}+\sigma_{\rm Y}$ & $f_{\ell}$  (mas yr$^{-1}$)\\
\noalign{\smallskip}\hline\noalign{\smallskip}
4 & $2.00\times 10^{-11}$ &  $8.1\times 10^{-12}$ & 7.4\\
6 & $2.92\times 10^{-11}$ &  $4.3\times 10^{-12}$ & 6.3\\
8 & $1.05\times 10^{-11}$ &  $3.0\times 10^{-12}$ & 0.4\\
10 & $7.8\times 10^{-12}$ &  $2.9\times 10^{-12}$ & 0.1\\
12 & $3.9\times 10^{-12}$ &  $1.8\times 10^{-12}$ & -\\
14 & $5\times 10^{-13}$ &  $1.7\times 10^{-12}$ & -\\
16 & $1.7\times 10^{-12}$ &  $1.4\times 10^{-12}$ & -\\
18 & $2\times 10^{-13}$ &  $1.4\times 10^{-12}$ & -\\
20 & $2.5\times 10^{-12}$ &  $1.4\times 10^{-12}$ & -\\

\noalign{\smallskip}\hline\noalign{\smallskip}
&    $\delta\mu = 30\%$ (SAV) & $\delta\mu = 20\%$ (RSS) &   \\  %
\noalign{\smallskip}\hline\noalign{\smallskip}
\end{tabular}
\end{table}
\begin{table}[!h]
   \caption{Bias due to the mismodelling in the even zonals of the models  X = JEM01-RL03B, based on 49 months of GRACE-only data, and Y = GGM03S \protect\cite{ggm03}.
    The $\sigma$ for GGM03S are calibrated. $\Delta \overline{C}_{\ell 0}$ are always larger than the linearly added sigmas apart from $\ell = 16$.}\label{tavolaJEM1}
\centering
\bigskip

\begin{tabular}{llll}
\hline\noalign{\smallskip}
$\ell$ & $\Delta\overline{C}_{\ell 0}$ (JEM01-RL03B-GGM03S) & $\sigma_{\rm  X}+\sigma_{\rm Y}$ & $f_{\ell}$  (mas yr$^{-1}$)\\
\noalign{\smallskip}\hline\noalign{\smallskip}
4 & $1.97\times 10^{-11}$ &  $4.3\times 10^{-12}$ & 7.3\\
6 & $2.7\times 10^{-12}$ &  $2.3\times 10^{-12}$ & 0.6\\
8 & $1.7\times 10^{-12}$ &  $1.6\times 10^{-12}$ & -\\
10 & $2.3\times 10^{-12}$ &  $8\times 10^{-13}$ & -\\
12 & $7\times 10^{-13}$ &  $7\times 10^{-13}$ & -\\
14 & $1.0\times 10^{-12}$ &  $6\times 10^{-13}$ & -\\
16 & $2\times 10^{-13}$ &  $5\times 10^{-13}$ & -\\
18 & $7\times 10^{-13}$ &  $5\times 10^{-13}$ & -\\
20 & $5\times 10^{-13}$ &  $4\times 10^{-13}$ & -\\

\noalign{\smallskip}\hline\noalign{\smallskip}
&    $\delta\mu = 17\%$ (SAV) & $\delta\mu = 15\%$ (RSS) &   \\  %
\noalign{\smallskip}\hline\noalign{\smallskip} %
\end{tabular}
\end{table}
\begin{table}[!h]
   \caption{Bias due to the mismodelling in the even zonals of the models  X = JEM01-RL03B and Y = ITG-Grace03s \protect\cite{itggrace03s}.
    The $\sigma$ for ITG-Grace03s are formal. $\Delta \overline{C}_{\ell 0}$ are always larger than the linearly added sigmas.}\label{tavolaJEM2}
\centering
\bigskip

\begin{tabular}{llll}
\hline\noalign{\smallskip}
$\ell$ & $\Delta\overline{C}_{\ell 0}$ (JEM01-RL03B-ITG-Grace03s) & $\sigma_{\rm  X}+\sigma_{\rm Y}$ & $f_{\ell}$  (mas yr$^{-1}$)\\
\noalign{\smallskip}\hline\noalign{\smallskip}
4 & $2.68\times 10^{-11}$ &  $4\times 10^{-13}$ & 9.9\\
6 & $3.0\times 10^{-12}$ &  $2\times 10^{-13}$ & 0.6\\
8 & $3.4\times 10^{-12}$ &  $1\times 10^{-13}$ & 0.1\\
10 & $3.6\times 10^{-12}$ &  $1\times 10^{-13}$ & -\\
12 & $6\times 10^{-13}$ &  $9\times 10^{-14}$ & -\\
14 & $1.7\times 10^{-12}$ &  $9\times 10^{-14}$ & -\\
16 & $4\times 10^{-13}$ &  $8\times 10^{-14}$ & -\\
18 & $4\times 10^{-13}$ &  $8\times 10^{-14}$ & -\\
20 & $7\times 10^{-13}$ &  $8\times 10^{-14}$ & -\\

\noalign{\smallskip}\hline\noalign{\smallskip}
&    $\delta\mu = 22\%$ (SAV) & $\delta\mu = 10\%$ (RSS) &   \\  %
\noalign{\smallskip}\hline\noalign{\smallskip}
\end{tabular}
\end{table}
\begin{table}[!h]
   \caption{Aliasing effect of the mismodelling in the even zonal harmonics estimated in the X=ITG-Grace03s \protect\cite{itggrace03s} and the Y=EIGEN-GRACE02S \protect\cite{eigengrace02s} models.  The covariance matrix $\sigma$ for ITG-Grace03s are formal, while the ones of EIGEN-GRACE02S are calibrated. $\Delta \overline{C}_{\ell 0}$ are larger than the linearly added sigmas for $\ell =4,...,20$, apart from $\ell=18$. }\label{tavola7}
\centering
\bigskip

\begin{tabular}{llll}
\hline\noalign{\smallskip}
$\ell$ & $\Delta\overline{C}_{\ell 0}$ (ITG-Grace03s-EIGEN-GRACE02S) & $\sigma_{\rm  X}+\sigma_{\rm Y}$ & $f_{\ell}$  (mas yr$^{-1}$)\\
\noalign{\smallskip}\hline\noalign{\smallskip}
4 & $2.72\times 10^{-11}$ &  $3.9\times 10^{-12}$ & 10.1\\
6 & $2.35\times 10^{-11}$ &  $2.0\times 10^{-12}$ & 5.1\\
8 & $1.23\times 10^{-11}$ &  $1.5\times 10^{-12}$ & 0.4\\
10 & $9.2\times 10^{-12}$ &  $2.1\times 10^{-12}$ & 0.1\\
12 & $4.1\times 10^{-12}$ &  $1.2\times 10^{-12}$ & -\\
14 & $5.8\times 10^{-12}$ &  $1.2\times 10^{-12}$ & -\\
16 & $3.4\times 10^{-12}$ &  $9\times 10^{-13}$ & -\\
18 & $5\times 10^{-13}$ &  $1.0\times 10^{-12}$ & -\\
20 & $1.8\times 10^{-12}$ &  $1.1\times 10^{-12}$ & -\\

\noalign{\smallskip}\hline\noalign{\smallskip}
&    $\delta\mu = 37\%$ (SAV) & $\delta\mu = 24\%$ (RSS) &   \\  %
\noalign{\smallskip}\hline\noalign{\smallskip} %
\end{tabular}

\end{table}
%
%
%
%
           %
 %
%
%
%
%
%
%
%
\begin{table}[!h]
   \caption{Bias due to the mismodelling in the even zonals of the models  X = JEM01-RL03B, based on 49 months of GRACE-only data, and Y = AIUB-GRACE01S \protect\cite{aiub}. The latter one was obtained from GPS satellite-to-satellite tracking data and K-band range-rate data out of the
period January 2003 to December 2003 using the Celestial Mechanics Approach.
No accelerometer data, no de-aliasing products, and no regularisation was
applied.
    The $\sigma$ for AIUB-GRACE01S are formal.
    $\Delta \overline{C}_{\ell 0}$ are always larger than the linearly added sigmas.
    }\label{tavolaAIUB1}
\centering
\bigskip

\begin{tabular}{llll}
\hline\noalign{\smallskip}
$\ell$ & $\Delta\overline{C}_{\ell 0}$ (JEM01-RL03B$-$AIUB-GRACE01S) & $\sigma_{\rm  X}+\sigma_{\rm Y}$ & $f_{\ell}$  (mas yr$^{-1}$)\\
\noalign{\smallskip}\hline\noalign{\smallskip}
4 & $2.95\times 10^{-11}$ &  $2.1\times 10^{-12}$ & 11\\
6 & $3.5\times 10^{-12}$  &  $1.3\times 10^{-12}$ & 0.8\\
8 & $2.14\times 10^{-11}$ &  $5\times 10^{-13}$ & 0.7\\
10 & $4.8\times 10^{-12}$ &  $5\times 10^{-13}$ & -\\
12 & $4.2\times 10^{-12}$ &  $5\times 10^{-13}$ & -\\
14 & $3.6\times 10^{-12}$ &  $5\times 10^{-13}$ & -\\
16 & $8\times 10^{-13}$ &    $5\times 10^{-13}$ & -\\
18 & $7\times 10^{-13}$  &    $5\times 10^{-13}$ & -\\
20 & $1.0\times 10^{-12}$ &    $5\times 10^{-13}$ & -\\

\noalign{\smallskip}\hline\noalign{\smallskip}
&   $\delta\mu = 26\%$ (SAV)&    $\delta\mu = 23\%$ (RSS)& \\  %
\noalign{\smallskip}\hline\noalign{\smallskip} %
\end{tabular}
\end{table}
\begin{table}[!h]
   \caption{Bias due to the mismodelling in the even zonals of the models  X = EIGEN-GRACE02S \protect\cite{eigengrace02s} and Y = AIUB-GRACE01S \protect\cite{aiub}. The $\sigma$ for AIUB-GRACE01S are formal, while those of EIGEN-GRACE02S are calibrated.
    $\Delta \overline{C}_{\ell 0}$ are  larger than the linearly added sigmas for $\ell=4,6,8,16$.
    }\label{tavolaAIUB2}
\centering
\bigskip

\begin{tabular}{llll}
\hline\noalign{\smallskip}
$\ell$ & $\Delta\overline{C}_{\ell 0}$ (EIGEN-GRACE02S$-$AIUB-GRACE01S) & $\sigma_{\rm  X}+\sigma_{\rm Y}$ & $f_{\ell}$  (mas yr$^{-1}$)\\
\noalign{\smallskip}\hline\noalign{\smallskip}
4 & $2.98\times 10^{-11}$ &  $6.0\times 10^{-12}$ & 11.1\\
6 & $2.29\times 10^{-11}$  &  $3.3\times 10^{-12}$ & 5.0\\
8 & $1.26\times 10^{-11}$ &  $1.9\times 10^{-12}$ & 0.4\\
10 & $6\times 10^{-13}$ &  $2.5\times 10^{-12}$ & -\\
12 & $5\times 10^{-13}$ &  $1.6\times 10^{-12}$ & -\\
14 & $5\times 10^{-13}$ &  $1.6\times 10^{-12}$ & -\\
16 & $2.9\times 10^{-12}$ &    $1.4\times 10^{-12}$ & -\\
18 & $6\times 10^{-13}$  &    $1.4\times 10^{-12}$ & -\\
20 & $2\times 10^{-13}$ &    $1.5\times 10^{-12}$ & -\\

\noalign{\smallskip}\hline\noalign{\smallskip}
&   $\delta\mu = 34\%$ (SAV)&    $\delta\mu = 25\%$ (RSS)& \\  %
\noalign{\smallskip}\hline\noalign{\smallskip} %
\end{tabular}
\end{table}

The systematic bias evaluated with a more realistic approach is about 3 to 4 times larger than one can obtain by only using this or that particular model. The scatter is still quite large and far from the $5-10\%$ claimed in Ref.~\cite{Ciu04}. In particular, it appears that $J_4$, $J_6$, and to a lesser extent $J_8$, which are just the most relevant zonals for us because of their impact on the combination of \rfr{combi}, are the most uncertain ones, with discrepancies $\Delta J_{\ell}$ between different models, in general, larger than the sum of their sigmas $\sigma_{J_{\ell}}$, calibrated or not.  

Another way to evaluate the uncertainty in the LAGEOS-LAGEOS II node test may consist of computing the nominal values of the total combined precessions for different models  and comparing them, i.e. by taking
\eqi \left|\sum_{\ell = 4}\left(\dot\Omega_{.\ell}^{\rm LAGEOS} + c_1\dot\Omega_{.\ell}^{\rm LAGEOS\ II}\right) [J_{\ell}(\rm  X)-J_{\ell}(\rm Y)]\right|.\eqf The results are shown in
Table \ref{tavolaa}.
 \begin{table}[!h]\caption{ Systematic uncertainty $\delta\mu$ in the LAGEOS-LAGEOS II test evaluated by taking the absolute value of the difference between the nominal values of the total combined node precessions due to the even zonals for different models X and Y, i.e. $\left|
 \dot\Omega^{\rm geopot}({\rm X})-\dot\Omega^{\rm geopot}({\rm Y})\right|$.}\label{tavolaa}
\centering
\bigskip

\begin{tabular}{ll}
\hline\noalign{\smallskip}
Models compared & $\delta\mu$  \\
\noalign{\smallskip}\hline\noalign{\smallskip}
AIUB-GRACE01S$-$JEM01-RL03B & $20\%$\\
%
%
AIUB-GRACE01S$-$GGM02S & $27\%$\\
AIUB-GRACE01S$-$GGM03S & $3\%$\\
AIUB-GRACE01S$-$ITG-Grace02 & $2\%$\\
AIUB-GRACE01S$-$ITG-Grace03 & $0.1\%$\\
%
%
AIUB-GRACE01S$-$EIGEN-GRACE02S & $33\%$\\
%
%
%
%
%
%
%
%
%
JEM01-RL03B$-$GGM02S & $7\%$\\
JEM01-RL03B$-$GGM03S & $17\%$\\
JEM01-RL03B$-$ITG-Grace02 & $18\%$\\
JEM01-RL03B$-$ITG-Grace03s & $20\%$\\
%
%
JEM01-RL03B$-$EIGEN-GRACE02S & $13\%$\\
%
%
GGM02S$-$GGM03S & $24\%$\\
GGM02S$-$ITG-Grace02& $25\%$\\
GGM02S$-$ITG-Grace03s& $27\%$\\
%
%
GGM02S$-$EIGEN-GRACE02S & $6\%$\\
%
%
GGM03S$-$ITG-Grace02 & $1\%$\\
GGM03S$-$ITG-Grace03s & $3\%$\\
%
%
GGM03S$-$EIGEN-GRACE02S & $30\%$\\
%
%
ITG-Grace02$-$ITG-Grace03s & $2\%$\\
%
%
ITG-Grace02$-$EIGEN-GRACE02S & $31\%$\\
%
%
%
ITG-Grace03s$-$EIGEN-GRACE02S & $33\%$\\
%
%
%
%
%
\noalign{\smallskip}\hline

\end{tabular}

\end{table}

A different approach that could be followed to take into account the scatter among the various solutions consists in computing mean and standard deviation of the entire set of values of the even zonals for the models considered so far, degree by degree, and    taking the standard deviations as representative of the uncertainties $\delta J_{\ell}, \ell = 4,6,8,...$. It yields $\delta\mu = 15\%$, in agreement with Ref.~\cite{Ries08}.

It must be recalled that also the further bias due to the cross-coupling between $J_2$ and the orbit inclination, evaluated to be about $9\%$ in Ref.~\cite{Ior07}, must be added.

\section{Possible alternative approaches to extract the Lense-Thirring signature from the data}\lb{approach}
The technique adopted so far in Ref.~\cite{Ciu04} and Ref.~\cite{Ries08} to extract the gravitomagentic signal from the LAGEOS and LAGEOS II data  is described in detail in, e.g.,
Refs.~\cite{LucBal06,Luc07}.
\begin{itemize}
\item In both the approaches the action of the gravitomagnetic field is not included in the dynamical force models used to fit the satellites' data. In the data reduction process no dedicated gravitomagnetic parameter is estimated, contrary to, e.g., station coordinates, state vector,
satellites' drag and radiation coefficients $C_D$ and $C_R$, respectively, etc.; the Lense-Thirring pattern is retrieved with a sort of post-post-fit analysis in which the time series of the computed\footnote{The expression ``residuals of the nodes'' is used, strictly speaking, in an improper sense because the Keplerian orbital elements are not directly measured.} ``residuals'' of the nodes with the difference between the orbital elements of consecutive arcs, combined with \rfr{combi}, is fitted with a straight line\footnote{Plus  some harmonic signals as well to remove the time-dependent tidal perturbations.}.


\item
In regard to possible other approaches which could be followed, it would be useful to, e.g., estimate (in the least square sense), among other solve-for parameters, purely phenomenological corrections $\Delta\dot\Omega$ to the LAGEOS/LAGEOS II node precessions as well, without modelling the Lense-Thirring itself, so that it will be, in principle, contained in $\Delta\dot\Omega$, and combine them according to \rfr{combi}.
Something similar has been done-although for different scopes-for the perihelia of the inner planets of the Solar System \cite{Pit05} and the periastron of the pulsars \cite{Kra06}.
To be more definite, various solutions with a complete suite of dynamical models, apart from the gravitomagnetic force itself, should be produced in which one inserts a further solve-for parameter, i.e. a correction  $\Delta\dot\Omega$ to the standard Newtonian modelled precessions.  One could see how the outcome varies by changing the data sets and/or the parameters to be solved for.
Maybe it could be done for each arc, so to have a collection of such node extra-rates. Such a strategy would be much more model-independent.

 \item Another previously suggested \cite{Nor01} way to tackle the problem consists of looking at a Lense-Thirring-dedicated parameter to be estimated along with all the zonals in a new global solution for the gravity field  incorporating the gravitomagnetic component as well; instead, in all the so far produced global gravity solutions no relativistic parameter(s) have been included in the set of the estimated ones.
\end{itemize}

A first, tentative step  towards the implementation of the strategy of the first point mentioned above with the LAGEOS satellites in terms  of the PPN parameter $\gamma$ has been recently taken by Combrinck in Ref.~\cite{Poz}.

\section{On the LARES mission}\lb{larez}
The combination that should be used for measuring the Lense-Thirring effect with LAGEOS, LAGEOS II and LARES is  \cite{IorNA}
\eqi \dot\Omega^{\rm LAGEOS}+k_1\dot\Omega^{\rm LAGEOS\ II}+ k_2\dot\Omega^{\rm LARES}.
\lb{combaz}\eqf
The coefficients $k_1$ and $k_2$ entering \rfr{combaz} are defined as
\begin{equation}
\begin{array}{lll}
k_1 = \rp{\cf 2{LARES}\cf4{LAGEOS}-\cf 2{LAGEOS}\cf 4{LARES}}{\cf 2{LAGEOS\ II}\cf 4{LARES}-\cf 2{LARES}\cf 4{LAGEOS\ II}}= 0.3586,\\\\
k_2 =  \rp{\cf 2{LAGEOS}\cf4{LAGEOS\ II}-\cf 2{LAGEOS\ II}\cf 4{LAGEOS}}{\cf 2{LAGEOS\ II}\cf 4{LARES}-\cf 2{LARES}\cf 4{LAGEOS\ II}}= 0.0751.
\end{array}\lb{cofis}
 \end{equation}
The combination of \rfr{combaz} cancels out, by construction, the impact of the first two even zonals; we have used $a_{\rm LR}=7828$ km, $i_{\rm LR}=71.5$ deg.
The total Lense-Thirring effect, according to \rfr{combaz} and \rfr{cofis}, amounts to 50.8 mas yr$^{-1}$.
\subsection{A conservative evaluation of the impact of the geopotential on the LARES mission}
The systematic error due to the uncancelled even zonals $J_6, J_8,...$ can be conservatively evaluated as
\eqi\delta\mu\leq \sum_{\ell = 6}\left|\dot\Omega^{\rm LAGEOS}_{.\ell}+k_1\dot\Omega^{\rm LAGEOS\ II}_{.\ell}+ k_2\dot\Omega^{\rm LARES}_{.\ell}\right|\delta J_{\ell}\lb{biass}\eqf

Of crucial importance is how to assess $\delta J_{\ell}$.  By proceeding as in Section \ref{grav} and by using the same models up to degree $\ell = 60$  because of the lower altitude of LARES with respect to LAGEOS and LAGEOS II which brings into play more even zonals, we have the results presented in Table \ref{tavolay}. They have been obtained with the standard and widely used Kaula approach \cite{Kau} in the following way. We, first, calibrated our numerical calculation with the analytical ones performed with the explicit expressions for $\dot\Omega_{.\ell}$ worked out up to $\ell=20$ in Ref.~\cite{Ior03}; then, after having obtained identical results, we confidently extended our numerical calculation to higher degrees by means of two different softwares \cite{IorioLARES1}.
 \begin{table}[!h]
 \caption{
 Systematic percent uncertainty $\delta\mu$ in the combined Lense-Thirring effect with LAGEOS, LAGEOS II and LARES according to \protect\rfr{biass} and
 $\delta J_{\ell}= \Delta J_{\ell}$
 up to degree
 $\ell = 60$
 for the global Earth's gravity solutions  considered here; the approach by Kaula \cite{Kau} has been followed. For LARES we adopted
 $a_{\rm LR}=7828$ km,
 $i_{\rm LR}=71.5$ deg, $e_{\rm LR} = 0.0$.}\label{tavolay}

\centering
\bigskip

\begin{tabular}{lll}
\hline\noalign{\smallskip}
Models compared ($\delta J_{\ell}=\Delta J_{\ell}$) & $\delta\mu$ (SAV) & $\delta\mu$ (RSS)\\
\noalign{\smallskip}\hline\noalign{\smallskip}
AIUB-GRACE01S$-$JEM01-RL03B & $23\%$ & $16\%$\\
%
%
AIUB-GRACE01S$-$GGM02S & $16\%$ & $8\%$\\
AIUB-GRACE01S$-$GGM03S & $22\%$ & $13\%$\\
AIUB-GRACE01S$-$ITG-Grace02 & $24\%$ & $15\%$\\
AIUB-GRACE01S$-$ITG-Grace03 & $22\%$ & $14\%$\\
%
%
AIUB-GRACE01S$-$EIGEN-GRACE02S & $14\%$ & $7\%$\\
%
%
%
%
%
%
%
%
%
JEM01-RL03B$-$GGM02S & $14\%$ & $9\%$  \\
JEM01-RL03B$-$GGM03S & $5\%$ & $3\%$  \\
JEM01-RL03B$-$ITG-Grace02 & $4\%$ & $2\%$  \\
JEM01-RL03B$-$ITG-Grace03s & $5\%$ & $2\%$  \\
%
%
JEM01-RL03B$-$EIGEN-GRACE02S & $26\%$ & $15\%$  \\
%
%
GGM02S$-$GGM03S & $13\%$ & $7\%$  \\
GGM02S$-$ITG-Grace02& $16\%$ & $8\%$  \\
GGM02S$-$ITG-Grace03s& $14\%$ & $7\%$  \\
%
%
GGM02S$-$EIGEN-GRACE02S & $14\%$ & $7\%$  \\
%
%
GGM03S$-$ITG-Grace02 & $3\%$ & $2\%$  \\
GGM03S$-$ITG-Grace03s & $2\%$ & $0.5\%$  \\
%
%
GGM03S$-$EIGEN-GRACE02S & $24\%$ & $13\%$  \\
%
%
ITG-Grace02$-$ITG-Grace03s & $3\%$ & $2\%$  \\
%
%
ITG-Grace02$-$EIGEN-GRACE02S & $25\%$ & $14\%$  \\
%
%
%
ITG-Grace03s$-$EIGEN-GRACE02S & $24\%$ & $13\%$  \\
%
%
%
%
%
\noalign{\smallskip}\hline

\end{tabular}

\end{table}

It must be stressed that they may be still optimistic: indeed, computations for $\ell > 60$ become unreliable because of numerical instability of the results.

In Table \ref{tavolax} we repeat the calculation by using for $\delta J_{\ell}$ the covariance matrix sigmas $\sigma_{J_{\ell}}$; also in this case we use the approach by Kaula \cite{Kau} up to degree $\ell = 60$.
 \begin{table}[!h]\caption{ Systematic percent uncertainty $\delta\mu$ in the combined Lense-Thirring effect with LAGEOS, LAGEOS II and LARES according to \protect\rfr{biass} and $\delta J_{\ell}= \sigma_{J_{\ell}}$ up to degree $\ell = 60$ for the global Earth's gravity solutions  considered here; the approach by Kaula \protect\cite{Kau} has been followed. For LARES we adopted $a_{\rm LR}=7828$ km, $i_{\rm LR}=71.5$ deg, $e_{\rm LR} = 0.0$.}\label{tavolax}
\centering
\bigskip

\begin{tabular}{lll}
\hline\noalign{\smallskip}
Model ($\delta J_{\ell}=\sigma_{\ell}$) & $\delta\mu$ (SAV) & $\delta\mu$ (RSS)\\
\noalign{\smallskip}\hline\noalign{\smallskip}%
AIUB-GRACE01S (formal) & $11\%$ & $9\%$\\
%
 %
JEM01-RL03B & $1\%$ & $0.9\%$\\
GGM03S (calibrated) & $5\%$ & $4\%$\\
GGM02S (formal) & $20\%$ & $15\%$\\
ITG-Grace03s (formal) & $0.3\%$ & $0.2\%$\\
ITG-Grace02s (formal) & $0.4\%$ & $0.2\%$\\
%
%
EIGEN-GRACE02S (calibrated) & $21\%$ & $17\%$ \\
 \noalign{\smallskip}\hline

\end{tabular}

\end{table}

If, instead, one assumes  $\delta J_{\ell}=s_{\ell},\ \ell=2,4,6,...$ i.e., the standard deviations of the sets of all the best estimates of $J_{\ell}$ for the models considered here the systematic bias, up to $\ell=60$, amounts to $12\%$ (SAV) and $6\%$ (RSS). Again, also this result may turn out to be optimistic for the same reasons as before; the analysis by Iorio of Ref.~\cite{IorioLARES2} seems to confirm such concerns.


It must be pointed out that the evaluations presented here rely upon calculations of the coefficients $\dot\Omega_{.\ell}$ performed with the well known standard approach by Kaula \cite{Kau}; it would be important to try to follow also different computational strategies in order to test them.
\subsection{The impact of some non-gravitational perturbations}
It is worthwhile noting that also the impact of the subtle non-gravitational perturbations will be different with respect to the original proposal because LARES will fly in a  lower orbit and its thermal behavior will  probably be different with respect to  LAGEOS and LAGEOS II.  The reduction of the impact of the thermal accelerations, like the Yarkovsky-Schach effects, should have  been reached with two concentric spheres. However, as explained in Ref.~\cite{Andres}, this solution will increase the floating potential of LARES because of the much higher electrical resistivity and, thus, the perturbative effects produced by the charged particle drag. Moreover, the atmospheric drag will increase also because of the lower orbit of the satellite, both in its neutral and charged components.  Indeed, although it does not affect directly the node $\Omega$, it induces a secular decrease of the inclination $i$ of a LAGEOS-like satellite \cite{Mil87} which translates into a further bias for the node itself according to
\eqi\delta\dot\Omega_{\rm drag}=\rp{3}{2}n\left(\rp{R}{a}\right)^2 \rp{\sin i\ J_2}{(1-e^2)^2}\delta i,\eqf in which $\delta i$ accounts not only for the measurement errors in the inclination, but also for any unmodelled/mismodelled dynamical effect on it. According to Ref.~\cite{Iordrag}, the secular decrease for LARES would amount to \eqi \left\langle\dert i t\right\rangle_{\rm LR}\approx -0.6\ {\rm mas}\ {\rm yr}^{-1}\eqf yielding a systematic uncertainty in the Lense-Thirring signal of \rfr{combaz} of about $3-9\%$ yr$^{-1}$. An analogous indirect node effect via the inclination could be induced by the thermal Yarkovski-Rubincam force as well \cite{Iordrag}.
Also the Earth's albedo, with its anisotropic components, may have a non-negligible  effect.

Let us point out the following issue as well. At present, it is not yet clear how the data of LAGEOS, LAGEOS II and LARES will be finally used by the proponent team in order to try to detect the Lense-Thirring effect. This could turn out to be a non-trivial matter because of the non-gravitational perturbations. Indeed, if, for instance, a combination\footnote{The impact of the geopotential is, by construction, unaffected with respect to the combination of \rfr{combaz}.}
\eqi\dot\Omega^{\rm LARES}+h_1\dot\Omega^{\rm LAGEOS}+ h_2\dot\Omega^{\rm LAGEOS\ II}\lb{altr}\eqf was adopted instead of that of \rfr{combaz}, the coefficients of the nodes of LAGEOS and LAGEOS II, in view of the lower altitude of LARES, would be
 \begin{equation}
\begin{array}{lll}
h_1 = \rp{\cf 2{LAGEOS\ II}\cf4{LARES}-\cf 2{LARES}\cf 4{LAGEOS\ II}}{\cf 2{LARES}\cf 4{LAGEOS\ II}-\cf 2{LAGEOS\ II}\cf 4{LAGEOS}}= 13.3215,\\\\
h_2 =  \rp{\cf 2{LARES}\cf4{LAGEOS}-\cf 2{LAGEOS}\cf 4{LARES}}{\cf 2{LAGEOS}\cf 4{LAGEOS\ II}-\cf 2{LAGEOS\ II}\cf 4{LAGEOS}}= 4.7744.
\end{array}\lb{cofcaz}
 \end{equation}
and the combined Lense-Thirring signal would amount to 676.8 mas yr$^{-1}$.
As a consequence, the direct and indirect effects of the non-gravitational\footnote{The same may hold also for time-dependent gravitational perturbations affecting the nodes of LAGEOS and LAGEOS II, like the tides.} perturbations on the nodes of LAGEOS and LAGEOS II would be enhanced by such larger coefficients and this may yield a degradation of the total obtainable accuracy.
\section{The Sun's Lense-Thirring effect on the inner planets}\lb{SolLT}
The action of the solar gravitomagnetic field on the
Mercury's longitude of perihelion\footnote{Since $\Omega$ and $\omega$ do not lie, in general, in the same plane, $\varpi$ is a \virg{dogleg} angle.} $\varpi\equiv\Omega+\omega$ was calculated for the first
time by De Sitter in Ref.~\cite{DeS} who, by assuming a homogenous and
uniformly rotating Sun, found a secular rate of $-0.01$
arcseconds per century (arcsec cy$^{-1}$ in the following). This value is
also quoted by Soffel in Ref.~\cite{Sof}; Cugusi and Proverbio \cite{Cug78}
yield $-0.02$ arcsec cy$^{-1}$  for the argument of perihelion $\omega$ of Mercury.

The recent estimate of the Sun's angular momentum \cite{Pij1,Pij2} from helioseismology yields a
precessional effect one order of magnitude smaller for Mercury; see Table \ref{chebolas} for the predicted Lense-Thirring precessions $\dot\varpi_{\rm LT}$ of the longitudes of the perihelia of the inner planets; they are of the order of $10^{-3}-10^{-5}$ arcsec cy$^{-1}$.  In computing them we have taken into account that the inclinations $I$ of the planets which are usually quoted in literature are referred to the mean ecliptic at a given epoch\footnote{It is J2000 (JD 2451545.0).}  \cite{Roy}, while the Sun's equator is tilted by $\epsilon_{\odot} = 7.15$ deg to the mean ecliptic at J2000 \cite{Carri}.
\begin{table}
\caption{First line: corrections ${\Delta\dot\varpi}$, in ${ 10^{-4} }$ {arcsec cy}${^{-1}}$ (1 arcsec cy$^{-1}=10$ mas yr$^{-1}=1.5\times 10^{-15}$ s$^{-1}$),  to the standard Newton/Einstein secular precessions of the perihelia
of the {inner} planets estimated by {E.V. Pitjeva} \protect\cite{Pit05} with the {EPM2004} ephemerides. The result for Venus (E.V. Pitjeva, private communication, 2008) has been obtained by recently processing radiometric
data from Magellan spacecraft with the EPM2008 ephemerides \protect\cite{PitX}. The errors in {brackets} are  the {\it formal}, {\it statistical} ones. Second line:  predicted Lense-Thirring perihelion precessions ${\dot\varpi_{\rm LT}}$, in ${ 10^{-4} }$ {arcsec cy}${^{-1}}$ \protect\cite{IorS1}.
Third line: nominal values of the perihelion precessions due to the Sun's oblateness for $J_2^{\odot}=2\times 10^{-7}$ \protect\cite{obloK,Pit05}; the current level of uncertainty in it is about $10\%$ \protect\cite{oblo2}. Fourth line: nominal values of the perihelion precessions due to the ring of the minor asteroids for $m_{\rm ring}=5\times 10^{-10}\ M_{\odot}$ \protect\cite{ring}; the uncertainty in it amounts to $\delta m_{\rm ring}=1\times 10^{-10}\ M_{\odot}$ \protect\cite{ring}. Fifth line: nominal values of the perihelion precessions due to a massive ring modelling the action of the Classical Kuiper Belt Objects (CKBOs) for $m_{\rm CKBOs}=0.052\ m_{\oplus}$ \protect\cite{ckbo}.\label{chebolas}
}
\centering
\bigskip
\begin{tabular}{@{}lllll@{}}
\noalign{\smallskip}\hline\noalign{\smallskip}
& {Mercury} & {Venus} & {Earth} & {Mars}  \\
 \noalign{\smallskip}\hline\noalign{\smallskip}
$\Delta\dot\varpi$ & $-36\pm 50\ (42)$ & $-4\pm 5\ (1)$ & $-2\pm 4\ (1)$ & $1\pm 5\ (1)$\\
\noalign{\smallskip}\hline\noalign{\smallskip}
$\dot\varpi_{\rm LT}$& $-20$ & $-3$  & $-1$ & $-0.3$\\
$\dot\varpi_{J_2^{\odot}}$ & $+254$ & $+26$ & $+8$ & $+2$\\
$\dot\varpi_{\rm  ring}$ & $+3$ & $+7$ & $+11$ & $+24$\\
$\dot\varpi_{\rm CKBOs}$ & $+0.2$ & $+0.6$ & $+1$ & $+2$\\
\noalign{\smallskip}\hline\noalign{\smallskip}

\end{tabular}
\end{table}
So far, the solar Lense-Thirring effect on the orbits of the inner planets
was believed to be too small to be detected \cite{Sof}.
However, the situation is now favorably changing. Iorio in Ref.~\cite{IorAA} preliminarily investigated the possibility of
measuring such tiny effects in view of recent important
developments in the planetary ephemerides generation.

First attempts to measure the Sun's Lense-Thirring effect have recently been
implemented by Iorio in Refs.~\cite{IorS1,IorS2} with the Ephemerides of Planets and the
Moon EPM2004 \cite{PitSS} produced by E.V. Pitjeva at the
Institute of Applied Astronomy (IAA) of the Russian
Academy of Sciences (RAS). They are based on a data set
of more than 317,000 observations (1913$-$2003) including
radiometric measurements of planets and spacecraft,
astrometric CCD observations of the outer planets and
their satellites, and meridian and photographic observations.
Such ephemerides were constructed by the simultaneous
numerical integration of the equations of motion for
all planets, the Sun, the Moon, 301 largest asteroids,
rotations of the Earth and the Moon, including the
perturbations from the solar quadrupolar mass moment
$J_2^{\odot}$ and asteroid ring that lies in the ecliptic plane and
consists of the remaining smaller asteroids. In regard to the
post-Newtonian dynamics, only the gravitoelectric, Schwarzschild-like terms of order  $\mathcal{O}(c^{-2})$,
in the harmonic gauge \cite{NewH}, were included; the gravitomagnetic force of the Sun was not modelled.

The EPM2004 ephemerides were used by Pitjeva in Ref.~\cite{Pit05} to phenomenologically
estimate corrections $\Delta\dot\varpi$ to the known standard Newtonian/Einsteinian secular
precessions of the longitudes of perihelia of the inner planets
as fitted parameters of a particular solution. In Table \ref{chebolas} it is
possible to find their values obtained by comparing the modelled
observations computed using the constructed ephemerides
with actual observations. In determining such
extra-precessions the PPN parameters  $\gamma$ and $\beta$
and the solar even zonal harmonic coefficient $J_2^{\odot}$ were not
fitted; they were held fixed to their general relativistic and Newtonian values, i.e.
$\gamma=\beta=1$, $J_2^{\odot} = 2\times 10^{-7}$. In the EPM2008 ephemerides \cite{PitX}, used to accurately estimate the perihelion precession of Venus as well, were constructed by also modelling the actions of Eris and of the other 20 largest Trans-Neptunian Objects (TNOs); the database was enlarged by including, among other things, ranging data to Magellan and Cassini spacecraft.

Although the original purpose\footnote{The goal of Pitjeva in Ref.~\cite{Pit05} was to make a test of the quality of the
previously obtained general solution in which certain values of $\beta,\gamma,J_2^{\odot}$
were obtained. If the construction of the ephemerides was satisfactory,
very small `residual' effects due to such parameters should have been
found. She writes: \virg{At present, as a test, we can determine [...] the
corrections to the motions of the planetary perihelia, which allows us to
judge whether the values of $\beta,\gamma$ and $J_2^{\odot}$ used to construct the ephemerides
are valid.}. The smallness of the extra-perihelion precessions found in her
particular test-solution is interpreted by Pitjeva as follows: \virg{Table 3 [of Ref.~\cite{Pit05}] shows
that the parameters $\gamma=\beta=1,J_2^{\odot}=2\times 10^{-7}$ used to construct the
EPM2004 ephemerides are in excellent agreement with the observations.}} of the estimation of the
corrections $\Delta\dot\varpi$ was not the measurement of the Lense-Thirring effect, the
results of Table \ref{chebolas} can be used to take first
steps towards an observational corroboration of the existence
of the solar gravitomagnetic force.

From Table \ref{chebolas} it turns out that the  magnitude of the Lense-Thirring perihelion precessions of the inner planets just lies at the edge of the accuracy in determining  $\Delta\dot\varpi$. All the predicted Lense-Thirring precessions are compatible with the estimated corrections $\Delta\dot\varpi$. Moreover, if one constructs the ratios $\mathcal{L}_{\rm AB}$ of the Lense-Thirring precessions \eqi{\mathcal{L}}_{\rm AB}\equiv\rp{ \dot\varpi^{\rm A}_{\rm LT}  }{ \dot\varpi^{\rm B}_{\rm LT}   }=\rp{   a^3_{\rm B}(1-e^2_{\rm B})^{3/2}   }{ a^3_{\rm A}(1-e^2_{\rm A})^{3/2} }\eqf  for all the pairs of inner planets A and B and compare them with the ratios \eqi\Pi_{\rm AB}\equiv\rp{ \Delta\dot\varpi_{\rm A} }{ \Delta\dot\varpi_{\rm B} }\eqf of the estimated corrections for the same pairs of planets A and B, it turns out that
\eqi  \Pi_{\rm AB} = {\mathcal{L}}_{\rm AB}\eqf  within the errors\footnote{In calculating the errors in $\Pi_{\rm AB}$ we linearly propagated the realistic errors in $\Delta\dot\varpi$ of Table \ref{chebolas}, not the formal ones.}.
It must be noted that for Venus the { formal}, statistical error in $\Delta\dot\varpi$ is smaller than the gravitomagnetic effect. Concerning the systematic alias due to the various competing dynamical effects listed in Table \ref{chebolas}, the present-day level of mismodelling in them would make them not particularly insidious (at $\approx 10\%$ level of accuracy), at least for Mercury and, especially, Venus, with the exception of the impact of the Sun's oblateness on Mercury; by the way, it would be possible to remove it by suitably designing a linear combination of the perihelia of Mercury and Venus, as done for the LAGEOS satellites. In principle, it would be possible to use also the nodes \cite{IorAA}, if only the corrections $\Delta\dot\Omega$ to their standard precessions were available. If and when other teams of astronomers will independently estimate their own corrections to the standard perihelion (and, hopefully, node as well) precessions with different ephemerides, it will be possible to fruitfully repeat the present test.
\section{The Lense-Thirring test in the gravitational field of Mars with the Mars Global Surveyor spacecraft}\lb{MarsLT}
 Iorio in Refs.~\cite{IorMGS,IorMGS2} proposed an interpretation of the time
series of the RMS orbit overlap differences \cite{Kon} of the out-of-plane part $\nu$
of the orbit of the Martian nearly polar artificial satellite Mars Global Surveyor
(MGS) over a time span $\Delta P$ of about 5 years (14 November
1999-14 January 2005 in \cite{IorMGS2})
in terms of the general relativistic
gravitomagnetic Lense-Thirring effect. It turned out that the average of such a time series over
$\Delta P$, normalized to the predicted Lense-Thirring
out-of-plane mean shift over the same time span, is $\mu=
1.0018\pm 0.0053$.

Our interpretation has recently been questioned by Krogh in Ref.~\cite{Krogh}.
The remarks concerning the analysis presented in Refs.~\cite{IorMGS,IorMGS2}
mainly deal with I) The observable used: Iorio in Refs.~\cite{IorMGS,IorMGS2} would have
misinterpreted the MGS data II) The confrontation between the
prediction of the gravitomagnetic Lense-Thirring shift and the
data over the chosen time span $\Delta P$: Iorio in Refs.~\cite{IorMGS,IorMGS2} would have
incorrectly compared the 1.6 m value of the out-of-plane average
orbit error released by Konopliv et al. in Ref.~\cite{Kon} for the entire MGS
data set to the Lense-Thirring
shift calculated for a shorter time interval $\Delta P$  III) The data set used: Iorio in Refs.~\cite{IorMGS,IorMGS2} discarded some of the initial months of the MGS data set
 IV) The bias--neglected by Iorio in Refs.~\cite{IorMGS,IorMGS2}--due to the multipolar expansion of the
Newtonian part of the martian gravity field, as pointed out in Refs.~\cite{Sin07,Fel07}
quoted in Ref.~\cite{Krogh} V) The impact of the atmospheric drag, neglected by Iorio in Refs.~\cite{IorMGS,IorMGS2}

Below we present our reply which, basically, consists of the following points \cite{CEJP1}. As further, independent tests,  here we present various linear fits to {\rm  different data sets} including, among others, the {\rm  full} time series of the {\rm  entire MGS data}  (4 February 1999--14 January 2005) as well; the predictions of general relativity turn out to be {\rm  always confirmed}. The analytical calculation of the competing aliasing effects due to both the gravitational and non-gravitational perturbations, which affect the {\rm  in-plane} orbital components of MGS, do {\rm  not} show up in the real data.
Moreover, the non-conservative forces, whose steadily refined modeling mainly improved the {\rm  in-plane} orbital components of MGS, {\rm  not} the {\rm  normal} one, exhibit high-frequency, non-cumulative in time variations.

%
\subsection{Our arguments}
\begin{itemize}
\item  [1)]
The entire MGS data  set  was subdivided by Konopliv et al. in Ref.~\cite{Kon} in 388 (not 442, as claimed by Krogh in Ref.~\cite{Krogh}) smaller time intervals
  of data called arcs. For MGS, the lengths of the arcs vary from 4 to
6 days, so to cover many orbital revolutions ($\approx 2$ h).
  For each arc, the spacecraft position and the velocity, among other things, were estimated and used as
  starting point for a numerical propagation of the satellite's motion by
  means of the dynamical models which, in the case of MGS, did not include the general relativistic
  gravitomagnetic force.
  Contiguous arcs were overlapped by an amount of just 2 h, i.e. one
  orbital revolution, and the RMS spacecraft position difference
  among the predicted positions propagated from the estimated ones in the previous arc
  and the estimated positions of the subsequent arc was computed. Since the arc overlaps cover just about one orbit,
  such RMS differences, in fact, account for any among measurement errors, random errors, systematic bias due to mismodeling/unmodeling dynamical forces
  yielding secular, i.e. averaged over one orbital
  revolution, effects, whatever their physical origin may be.
  Indeed, RMS of orbit solution overlaps are {\rm  commonly used in satellite
  geodesy as useful and significant indicators of the
  overall orbit accuracy} \cite{Tap,LucBal06}.
  Conversely, they are also used to gain
  information about {\rm  systematic errors coming from inaccurate
  modeling of the forces acting on the spacecraft}. For details
  see Refs.~\cite{Tap,LucBal06}.
  Of course, such a technique is insensitive to
  short-period effects, i.e. having frequencies
  higher than the orbital one: only dynamical features of
  motion with time-scales equal to, or larger than one orbital period
  can be sensed by such orbit overlap differences. Moreover, the
  average orbit error $\left\langle\Delta \nu_{\rm
diff}\right\rangle$ of about 1.6 m does not refer to this or that
  particular arc overlap; instead, it comes from the mean
  of the entire set of RMS orbit overlap differences for the chosen time span $\Delta
  P$ and is well representative of those un-modelled/mis-modelled
  forces yielding effects which do not average out over $\Delta
  P$, as it is just the case of the Lense-Thirring signal.
  Time-varying patterns exhibiting well-defined periodicities-including  also measurement errors like, e.g., those related to
   the Earth-Mars geometry-are, instead,
  mainly averaged out yielding little or no contribution to
  the average orbit error. Incidentally, from the above discussion about
  the meaning of the average orbit error, it should be apparent that it does
  not make sense to look for the error of the error, as, instead, seemingly
  required by Krogh in Ref.~\cite{Krogh} when he blames Iorio \cite{IorMGS2} for not having included the uncertainty in $\left\langle\Delta \nu_{\rm
res}\right\rangle$. Another criticism in Ref.~\cite{Krogh} is that the
RMS overlap differences would be unable to specify any orbital precession.

To reply to all such criticisms we decide to perform another, independent test of our hypothesis.
First, by linearly
fitting\footnote{Note that, since the plots in Fig. 3 of Ref.~\cite{Kon} are
{\rm  semi-logarithmic}, one should {\rm  not} visually look for a straight line
in them.} the {\rm  full time series}  of Ref.~\cite{Kon}, after having rescaled the data points in order to shift the zero point of the time-series to the middle of the data span, we get a slope of $-0.64\pm 0.26$ m yr$^{-1}$, (with 95$\%$ confidence bounds), while the predicted
Lense-Thirring MGS out-of-plane rate (customarily defined {\rm  positive}
{\rm  along} the spacecraft's orbital angular momentum) amounts to 0.62 m yr$^{-1}$.
The obtained minus sign is due to the fact that Konopliv et al. in Ref.~\cite{Kon}
defined the normal direction to be positive in the {\rm  opposite} direction of
the MGS orbital angular momentum (Konopliv 2007, private communication). Should such a linear fit be used
as indicator of the existence of the Lense-Thirring effect, its relativistic prediction would be fully confirmed
within the experimental error; instead,
the hypothesis of a null effect would be rejected at 2.4 sigma level.
Then, we also repeat our procedure by fitting with a straight line the entire data set {\rm  without full January 2001}, mainly affected by likely measurement errors which, according to Krogh \cite{Krogh}, would mimic the Lense-Thirring effect, getting  $-0.61\pm 0.26$ m yr$^{-1}$.  {\rm  The removal of the entire year 2001}, mainly affected by angular momentum wheel desaturation operations, yields $-0.57\pm 0.28$ m yr$^{-1}$. Another linear fit to the time series after removing the last month (December 2004-January 2005) yields $-0.62\pm 0.27$ m yr$^{-1}$.

Such results reply to the criticisms II) and III) as well concerning $\Delta P$, to which a large part of Ref.~\cite{Krogh} is devoted.

             \item [IV)] Krogh \cite{Krogh} quotes  Ref.~\cite{Sin07}  in which analytical calculation about the corrupting impact of various physical parameters of Mars through the classical node precessions induced by the even zonal harmonic coefficients $J_{\ell}$ of the multipolar expansion of the Newtonian part of the martian gravitational potential are presented.   In particular,
                 the authors of Ref.~\cite{Sin07}  use the first five even zonals $J_2...J_{10}$ along with their associated errors from former global solutions for the Mars'gravity field, the uncertainty in the Mars'  $GM$ and in the MGS semimajor axis and inclination, plug them into analytical formulas for the classical secular node precessions and conclude that, since the resulting effect is tens of thousand times larger than the Lense-Thirring effect on MGS, this would be fatal for any attempt to detect the gravitomagnetic frame-dragging with such a spacecraft.

                 The point is that {\rm  such figures}, as others which can be obtained from more accurate calculation, {\rm  must ultimately be compared with the reality of the data}, i.e. the RMS orbit overlap differences of MGS.

                 We, in fact, repeated such calculation by considering also the other even zonals up to $J_{20}$ along with the latest errors of the  MGS95J global solution and including the uncertainties in the Mars' radius as well. By summing, in a root-sum-square fashion, such  terms we get a mean bias of 78.9 m d$^{-1}$ in the out-of-plane MGS orbital component: by linearly summing them we get an upper bound of $111.6$  m d$^{-1}$.
                 Such figures clearly show how they are by far not representative of the real MGS orbit. Indeed, over a time span of 5 years we would have an {\rm  enormous mean shift as large as 144 km} (root-sum-square calculation) or {\rm  203 km} (linear sum). Interestingly, even if the set of the RMS overlap differences of MGS were to be considered as representative of a single orbital arc 6 d long only, the conclusion would be the same: indeed, in {\rm  this} case, the total cross-track mean shift due to the martian gravitational potential {\rm  would amount to 473.1 m} (root-sum-square) or {\rm  669.6 m} (linear sum).

                 In regard to Ref.~\cite{Fel07}, quoted in Ref.~\cite{Krogh} as well, let us recall again that the RMS orbit overlap differences are just used to account, in general, for all the measuremnt/systematic errors giving an indication of the overall orbit accuracy
                 \cite{Tap,LucBal06}. The important point is that {\rm  they cancel out, by construction,} errors, systematic or not, {\rm  common to consecutive arcs}$-$it would just be the case of a bias like that described in Ref.~\cite{Fel07}$-$, while effects like the Lense-Thirring one, accumulating in time, are, instead, singled out \cite{LucBal06}.

                  \item [V)] In regard to the impact of the non-gravitational perturbations,   the authors of Ref.~\cite{Sin07}
yield a total un-modelled  non-gravitational acceleration of
$\approx 10^{-11}$ m s$^{-2}$ which is the same order of magnitude
of the Lense-Thirring acceleration induced by Mars on MGS. They  neither present any detailed calculation of the effect of such an
acceleration on the normal portion of the MGS orbit nor specify if such a magnitude refers to the out-of-plane component. However, some
simple considerations can be easily traced: a hypothetic, generic perturbing out-of-plane
force 6.7 times larger than the Lense-Thirring one and having the same time signature, i.e. linear in time, should
induce a 10.8 m cross-track shift, on average, over the considered
time span $\Delta P$. Again, {\rm  such a bias is neatly absent from the data}.
 By the way, as clearly stated in \cite{Kon},
it is the {\rm  along-track} portion of the MGS orbit$-$left unaffected
by the Lense-Thirring force$-${\rm  to be mainly perturbed by the
non-gravitational forces}: indeed, the along-track empirical
accelerations fitted in Ref.~\cite{Kon} amount just to
$\approx 10^{-11}$ m s$^{-2}$, which shows that the guess by \cite{Sin07} is somewhat correct, but it refers to the {\rm  along-track} component.

Time-dependent, periodic signatures would, instead, be averaged
out, provided that their characteristic time scales are relatively short, as it is just the case. Indeed, the non-conservative accelerations, which are especially active in the MGS {\rm  in-plane} orbital components as clearly stated in Refs.~\cite{Lem01,Kon}, exhibit time-varying patterns over 12 hr \cite{Lem01} which, hypothetically mapped to the out-of-plane direction, are averaged
out over  multi-year time spans  (and, incidentally, over 6 d as well).
To be more definite,
in regard to the issue of the impact of the atmospheric drag on the
  cross-track portion of the orbit of MGS, raised in  Ref.~\cite{Krogh}, let us note that
  it requires not only  to consider the node $\Omega$, as apparently claimed by Krogh \cite{Krogh}, but also the inclination $i$ according to Ref.~\cite{Cri}
  \eqi\Delta \nu=a\sqrt{\left(1+\rp{e^2}{2}\right)\left[\rp{(\Delta i)^2}{2} + (\sin i\Delta\Omega)^2\right]}.\lb{tra}\eqf
  According to, e.g., Ref.~\cite{Mil87}, the perturbing acceleration $\dr$ due to the atmospheric drag
  can be cast into the form
  \eqi \dr=-\rp{1}{2}ZC_{\rm D}\rp{\Sigma}{M}\rho v\bds v,\eqf
  where $\Sigma/M$ is the spacecraft cross sectional area (perpendicular to the
  velocity) divided by its mass, $C_{\rm D}$ is the drag
  coefficient, $\rho$ is the atmospheric  density (assumed to be constant over one orbital
  revolution), $\bds v$ is the satellite velocity in a
  planetocentric, non-rotating frame of reference and $Z$ is a
  corrective coefficient accounting for the fact  that the
  atmosphere is not at rest, but rotates with angular velocity $\psi_{\rm A}$
  more or less rigidly with
  the planet; $Z\approx 1$ for polar orbits \cite{Mil87}.
  While the secular, i.e. averaged over one orbital
  period $T$, drag shift on the node vanishes, it is not so for
  the inclination: indeed, it turns out \cite{Mil87}
  \eqi \left\langle\Delta i\right\rangle_{T} \approx \pi\left(\rp{A_{\rm drag}}{n^2 a}\right)\rp{\psi_{\rm A}}{n}
  +{\mathcal{O}}(e). \eqf As a result, the orbital plane tends to approach
  the planet's equator; the terms in brackets is the ratio of the
  drag force to the Newtonian monopole. As usual in perturbation theory, $a$ is meant as evaluated on the unperturbed reference
  ellipse. Thus, the out-of-plane drag shift
  is from \rfr{tra}
  \eqi \left\langle\Delta \nu_{\rm drag}\right\rangle\approx a\rp{\left\langle\Delta i_{\rm drag}\right\rangle}{\sqrt{2}}.\eqf
  In the following we will assume that $\psi_{\rm A}\approx\psi_{\rm Mars}=7.10\times 10^{-5}$ s$^{-1}$.
  Let us see what happens in the (unlikely) worst-case $A_{\rm drag}$  $\approx 10^{-11}$ m s$^{-2}$; it turns out that \eqi \left\langle\Delta
  \nu_{\rm drag}\right\rangle_{T}\sim 1\times 10^{-5}\ {\rm m}.\lb{drr}\eqf But $A_{\rm drag}$ is not constant over time spans days or years long \cite{For06}, so that such an effect is not a concern here.  By the way, even if it was not so, by assuming a $\approx 10\%$ mismodeling in
  drag--which is, in fact, modeled in Ref.~\cite{Kon}--\rfr{drr}, mapped onto about 5 yr, would give a $\approx 0.7\%$ uncertainty.

  Finally, Krogh \cite{Krogh} remarks that decreasing in the averages of the RMS orbit overlaps occurred in view of constantly improved modeling \cite{Lem01,Yua01}, but he does not recognize that the better modeling of the non-gravitational forces acting on MGS introduced in
  Ref.~\cite{Kon}
with respect to
previous works \cite{Lem01,Yua01} in which the Lense-Thirring effect was not modelled as well,
{\rm  only affected in a relevant way just the along-track} RMS overlap differences (a factor 10 better than in Refs.~\cite{Lem01,Yua01}),
{\rm  not the normal ones} (just a factor 2 better than in Refs.~\cite{Lem01,Yua01}).
                                             Moreover,
if the relativistic signature was removed or not present at all so that the determined
out-of-plane RMS overlap differences were only (or mainly) due to other causes
like mismodeling or unmodeling in the non-gravitational forces,
it is difficult to understand why the along-track RMS overlap differences
(middle panel of Figure 3 of Ref.~\cite{Kon}) have almost the same
magnitude, since the along-track component of the MGS orbit is
much more affected by the non-gravitational accelerations (e.g.
the atmospheric drag) than the out-of-plane one.
\end{itemize}

\section{Conclusions}
\begin{itemize}
  \item The present and future tests of the Lense-Thirring effect in the gravitational field of the Earth with some existing and future SLR artificial satellites are challenged by several competing classical forces whose impact is difficult to be assessed in a realistic and conservative way.
  \begin{itemize}
          \item Because of the lingering statistically significant discrepancies among the  estimated values of the low-degree even zonal harmonics of the geopotential in several GRACE-based global solutions produced by different institutions worldwide, the present-day uncertainty in the Lense-Thirring test performed by linearly combining the nodes of the  LAGEOS and LAGEOS II terrestrial satellites is larger than the claimed value of $5-10\%$ by a factor as large as up to 2-3 times.
          \item A different methodology to extract the Lense-Thirring signal from the  data of the LAGEOS-type satellites should be used to make a complementary test which would enforce (or disprove) those performed so far. That is, the gravitomagnetic force should be explicitly included in the dynamical force models of  LAGEOS and LAGEOS II, and a dedicated parameter accounting for it should be estimated in the least-square fitting of the observations of the LAGEOS satellites; moreover,  the changes, if any, in the values of the other estimated parameters with respect to the case in which the gravitomagnetic force was not modelled may be inspected as well. Alternatively, corrections $\Delta\dot\Omega$ to the standard Newtonian node precessions of the LAGEOS satellites may be estimated in a purely phenomenological way without modelling the Lense-Thirring effect; their values should, then, be compared with the predicted gravitomagnetic node precessions. An analogous procedure was followed with the periastron rates in the binary pulsar systems and with the perihelion precessions of the inner planets of the Solar System, although for different scopes.  Finally, another feasible approach, at least in principle, consists of producing a new global gravity solution by re-processing  the GRACE observations with dynamical force models including the gravitomagnetic force as well, and estimating it along with all the even zonals of the geopotential and inspecting the resulting covariance matrix.
          \item The LAGEOS-type LARES satellite, approved by ASI, should be launched at the end of 2009 or at the beginning of 2010 with a VEGA rocket: its  goal is a measurement of the Lense-Thirring effect with an accuracy of the order of $1\%$. Unfortunately, its orbital configuration will be different from the originally proposed one, with an altitude of about 1400 km; this may pose serious problems in terms of the impact of the even zonal harmonics of the geopotential because a much larger number of them should come into play by degrading the total accuracy obtainable. Extensive calculations performed with standard geodetic techniques and several global gravity model solutions point towards a systematic uncertainty which may be orders of magnitude larger than the claimed $1\%$. Also certain subtle issues related to the indirect impact of the atmospheric drag on the node precessions may contribute to further increasing the total error.
        \end{itemize}
  \item Recent improvements in the field of the planetary ephemerides make the possibility of measuring the Lense-Thirring effect in the gravitational field of the Sun with the inner planets much more promising than in the past. At present, the magnitude of the  Lense-Thirring perihelion precessions, not included in the models of the dynamical theories with which the planetary observations are reduced, lies just at the edge of the accuracy in determining the secular perihelion precessions of the rocky planets. It turns out that the predicted gravitomagnetic rates are in agreement with the recently estimated extra-precessions of the perihelia for all the inner planets; the same also holds  for the ratios of the perihelia for all the pairs of inner planets. Concerning the systematic errors which may be caused by the imperfect knowledge of several competing forces of classical origin, the most insidious one is the Sun's oblateness. However, its impact can be removed by linearly combining the perihelia of the planets as done for the LAGEOS-LAGEOS II test. The best suited candidates are Mercury and Venus whose orbits should be determined with an higher level of accuracy in the near future when ranging data to present and future planned spacecraft orbiting them will be collected and processed.  It would also be important if  other teams of astronomers independently would estimate their own corrections $\Delta\dot\varpi$ to the usual perihelion (and, hopefully, node as well) precessions of the inner planets  with different ephemerides to repeat the tests discussed here.
  \item Another Solar System scenario which could also be used for testing the Lense-Thirring effect is the gravitational field of Mars. In this framework, we recently proposed an interpretation of some orbital data of the nearly polar Mars Global Surveyor probe in terms of the gravitomagnetic force of Mars by performing two independent preliminary tests.
      The  {average} $ {\left\langle\Delta\nu\right\rangle}$ of the  {time series} of the out-of-plane RMS orbit  {overlap differences} of MGS, {normalized} to the  {average Lense-Thirring shift} $ {\left\langle\Delta\nu_{\rm LT}\right\rangle}$ over the same time span, is $ {\mu = 1.002\pm 0.005}$.
   {Linear fits} of the  {time series} for  {different time spans}, i.e. including different sets of data points, yield results  {in agreement} with the  {LT interpretation}. For example, the fit of the  {entire} data set yields a slope of  $ {1.03\pm 0.42}$, in  {LT-normalized units}; {removing} the data for the month  {January 2001}, likely affected by  {major measurement errors}, yields a normalized slope $ {0.98\pm 0.42}$.

\end{itemize}

\acknowledgments{I gratefully thank the organizers and the entire staff of this prestigious and high-quality international school for their kind invitation, their exquisite hospitality and  the financial support received.}

\end{document}